# Collisional cooling of light ions by co-trapped heavy atoms

Sourav Dutta, Rahul Sawant, and S. A. Rangwala

*Raman Research Institute, C. V. Raman Avenue, Sadashivanagar, Bangalore 560080, India*

We experimentally demonstrate cooling of trapped ions by collisions with co-trapped, higher mass neutral atoms. It is shown that the lighter $^{39}$K$^+$ ions, created by ionizing $^{39}$K atoms in a magneto-optical trap (MOT), when trapped in an ion trap and subsequently allowed to cool by collisions with ultracold, heavier $^{85}$Rb atoms in a MOT, exhibit a longer trap lifetime than without the localized $^{85}$Rb MOT atoms. A similar cooling of trapped $^{85}$Rb$^+$ ions by ultracold $^{133}$Cs atoms in a MOT is also demonstrated in a different experimental configuration to validate this mechanism of ion cooling by localized and centered ultracold neutral atoms. Our results suggest that cooling of ions by localized cold atoms holds for any mass ratio, thereby enabling studies on a wider class of atom-ion systems irrespective of their masses.

The cooling and trapping of dilute gases of ions [1,2] and atoms [3] has led to unprecedented precision in spectroscopy [4] and the study of interacting many particle systems [5]. Co-trapping ions and atoms [6–14] widens the scope of inquiry to the two-particle $1/r^4$ asymptotic interaction. Among the different methods to cool trapped ions, cooling by elastic collisions with cold neutral atoms is arguably the most generic. Indeed, this has been extensively used in buffer gas cooling of relatively heavy ions by collisions with cold lower mass neutral atoms. However, the complementary phenomenon, that of cooling of low-mass ions by collisions with heavier neutral atoms, has never been demonstrated experimentally. This is partly because calculations show that trapped ions in a uniform buffer gas will heat up when the ratio of the atom mass ($m_a$) to the ion mass ($m_i$) exceeds a critical value [15–18]. In recent years, theoretical studies suggest a possible experimental resolution by using a localized ensemble of ultracold neutral atoms placed precisely at the centre of the ion trap [12,19].

In this Letter, we experimentally demonstrate the cooling of low mass $^{39}$K$^+$ ions in a Paul trap by localized, heavier $^{85}$Rb atoms in a magneto optical trap (MOT) and validate our experimental results with numerical simulations. We also demonstrate cooling of trapped $^{85}$Rb$^+$ ions by ultracold $^{133}$Cs atoms in a MOT. The cooling manifests itself through an increase in the lifetime of the trapped ions. Our results support the possibility that cooling of ions by localized cold atoms may hold for any mass ratio thereby enabling studies on the other half of ultracold atom-ion systems irrespective of their masses.

The atom-ion mass ratios ($m_a/m_i = 2.179$ for $^{85}$Rb-$^{39}$K$^+$ and = 1.565 for $^{133}$Cs-$^{85}$Rb$^+$) we demonstrate cooling for, exceed the critical mass ratios (CMRs) [15–18] beyond which ion heating is predicted for uniform atomic gas densities. The experiment is consistent with our Monte Carlo (MC) simulations and other theoretical models [19] that consider collisions with centrally localized density (as opposed to uniform density) of cold atoms. We further demonstrate the competing ion heating mechanism due to collisions with the background gas vapour. The dependence of steady state ion temperature on the size of the cold atomic cloud is also established numerically. The present work lays the foundation for studies on a wider class of sympathetically cooled ion-atom mixtures and the resulting clarity may enable the realization of the few-partial-wave regime.

*Trapping and cooling ions and atoms.* The conditions and mechanisms for trapping atoms and ions are different because the atom is neutral and the ion charged. Atom traps typically have small trap depths and use a combination of static magnetic and/or optical fields. Ions on the other hand are trapped in Penning or Paul traps [20], of which the latter, a dynamic trap with time varying electric fields, is compatible [15] with overlapped cold atom ensembles because the perturbation of the Paul trap operation on co-trapped cold atoms is negligible [6,21]. It is for this reason that hybrid traps combine Paul traps for ions [1] with atom traps such as a MOT [7,10–13], magnetic trap [8] or optical dipole trap [8,9,22]. However, due to the dynamic trapping of ions in a Paul trap [1,20], the trapped ions in absence of laser cooling are more energetic than the atoms [12,13].

*Signature of cooling of trapped ions.* The motion of the dynamically trapped ions is described theoretically by the Mathieu equations, which are parameterized in terms of dimensionless $a$ and $q$ parameters, the values of which characterize the ion trap operation. The ion motion is separable [1] into (*i*) the micromotion, which is synchronous with the time varying radiofrequency (rf) field and (*ii*) the secular motion (macromotion), which is the low frequency oscillation of the ion in the effective trapping potential that the ions experience in the dynamic trapping fields. The micromotion is position dependent while the secular motion determines the size of the trapped ion's orbit and characterizes the trapped ion's temperature. The smaller the trapped ion's orbit the lower the ion's temperature, and therefore the task of cooling the ion is reduced to arresting the amplitude of the secular motion.

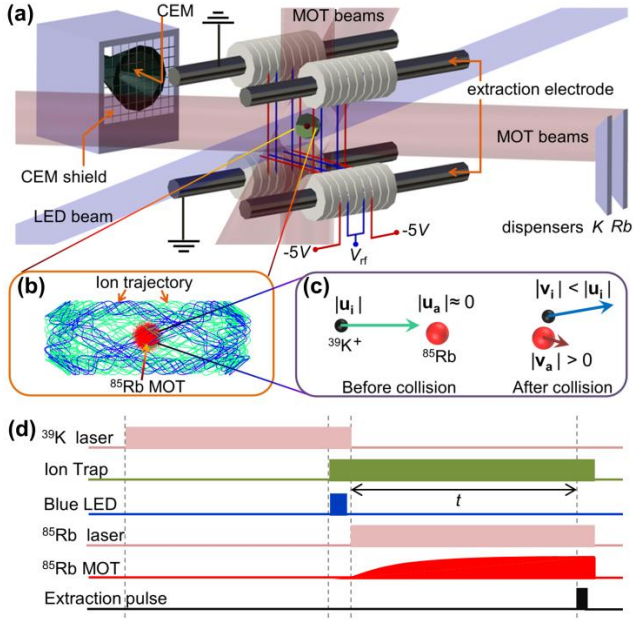

FIG. 1 (Color online). (a) A schematic of the experimental set-up. The rf voltage is fed to the two central wire electrodes and the two external wire electrodes serve as the end caps. The extraction of the trapped ions to the CEM is done by biasing two of the rod extraction electrodes appropriately after the hold time. (b) Detail of overlap volumes of the $^{85}$Rb MOT and a calculated, representative ion trajectory. (c) Illustrative diagram for transfer of momentum from the $^{39}$K$^+$ ion to the almost stationary $^{85}$Rb MOT atom, resulting in a lower kinetic energy of the ion post collision and a hot atom which usually escapes the MOT. (d) The timing sequence for the experiment.

*Criteria for ion heating and cooling.* The pioneering work on collisional cooling of ions by neutral atoms, by Major and Dehmelt [15], considers the ion trap volume within a constant density of buffer gas atoms. In this case it was found that ions collisionally heat when $m_a > m_i$, cool when $m_a < m_i$ and do not change temperature when $m_a = m_i$. The ratio $m_a/m_i = 1$, is then the theoretical CMR differentiating the cooling and heating regimes. In practice, the cooling by buffer gas has found extensive applications when $m_a \ll m_i$ [23].

In recent years, several independent theoretical analyses for ion cooling by neutral atoms have arrived at limiting CMRs [16–18] that are different from 1 (CMR = 1.55 [16], = 1.47 [17], = 0.95 [18] for our K$^+$ ion trap parameters).

Exceeding these CMRs results in the heating of ions (and below these cooling occurs) if the ion is assumed to be located within a cold gas of uniform density. However, in hybrid ion-atom traps the atomic density is localized, for example in a MOT, and is located at the centre of the ion trap [see Figs. 1(a) and 1(b)]. When this localized and precisely centred nature of the atomic density profile is taken into account, the trapped ions are expected to be cooled by elastic collisions with the ultracold atoms irrespective of the atom-ion mass ratio [12,17,19]. At the centre of the ion trap, the ion's micromotion is negligible while the ion's secular speed is the greatest – thus a collision with an ultracold atom, that is essentially at rest, always results in reduction in the ion's secular motion and hence in cooling of the ion [see Figs. 1(b) and 1(c)].

*Experimental procedure.* The experimental apparatus is described over several articles [24–26], where the thin wire modified spherical Paul trap's principles, operation and performance is characterized. The schematic diagram of the experimental apparatus is illustrated in Figs. 1(a)-(c). The ion trap has a depth $U \approx 0.5$ eV, $q = 0.42$ and $a \approx 0$. For this experiment, we have a vapour loaded dual MOT of $^{39}$K and $^{85}$Rb in operation. The $^{39}$K and $^{85}$Rb MOTs are very well overlapped with the ion trap centre and so the ions created from the MOT are efficiently loaded into the ion trap. The loading of the two MOTs are independently controlled by mechanical shutters placed in the paths of the respective laser beams. The $^{39}$K$^+$ ions are created by resonant two photon ionization of atoms in the $^{39}$K MOT, where the first photon (cooling laser) results in the $4S_{1/2} \rightarrow 4P_{3/2}$ excitation at 767 nm and a second photon is sourced from a collimated light emitting diode (LED) with emission centred at 456 nm. Further experimental details are provided in the Supplementary Material (SM) [27].

The experiment measures the lifetime of the trapped $^{39}$K$^+$ ions when held with and without the $^{85}$Rb MOT atoms. The experimental sequence is illustrated in Fig. 1(d). As the LED light can also ionize $^{85}$Rb MOT atoms in a similar two photon process, the MOTs are operated sequentially to prevent the ionization of $^{85}$Rb atoms. First, the $^{39}$K MOT is loaded, ion trapping fields are switched on, $^{39}$K$^+$ ions are created by switching on the blue LED briefly, and the $^{39}$K MOT light is shuttered off which empties the $^{39}$K MOT. Subsequently, either the $^{85}$Rb MOT light is allowed in and the $^{85}$Rb MOT loaded or the $^{85}$Rb light is kept blocked and no MOT is loaded. The trapped $^{39}$K$^+$ ions are held for variable hold times and extracted for detection by a channel electron multiplier (CEM), where individual ion arrivals register as individual ~8 ns pulses [24,25]. These pulses are counted and the surviving ion number in the ion trap determined as a function of ion hold time ($t$), in the presence or absence of the $^{85}$Rb MOT.

*Potassium ion cooling by rubidium atoms.* In Fig. 2(a), we present a representative plot (in semi-log scale) showing that a larger number of $^{39}$K$^+$ ions survive when held in the presence of the $^{85}$Rb MOT. As the $^{85}$Rb MOT takes ~ 5 s to load, initially there is no significant difference between the number of $^{39}$K$^+$ ions counted when $^{85}$Rb MOT is off (square) or on (circle), but by 25 s a clear separation in the number of surviving ions emerges. The higher number of trapped $^{39}$K$^+$ ions observed in the presence of the $^{85}$Rb MOT is due to the larger survival probability of the trapped ions in the presence of $^{85}$Rb MOT atoms. The larger survival probability of the ions can only be explained by the cooling of the trapped ions by the $^{85}$Rb MOT atoms [12,13]. If instead the ions were heating, the survival probability for $^{39}$K$^+$ ions would be lower when in contact with $^{85}$Rb MOT atoms. The experiments were repeated with $^{40}$K$^+$ ions instead of $^{39}$K$^+$ ions, and cooling of $^{40}$K$^+$ ions by $^{85}$Rb atoms was also observed. The cooling of $^{39}$K$^+$ ions occurs because the ultracold $^{85}$Rb atoms are localized precisely at the centre of the ion trap, and $V_a \ll V_i$, where $V_a$ ($V_i$) is the volume of the trapped atoms (ions).

The experiment of Fig. 2(a) is repeated nine times and similar plots are obtained. In each case, the ion decay is relatively fast up to $t \sim 5$ s and two cases, with and without MOT, are indistinguishable, consistent with the $^{85}$Rb MOT loading time. For $t \geq 5$ s, the number of ions $N(t)$ in both cases decays exponentially but with different lifetimes. We fit a single exponential decay of the form $N(t) = N_i\, e^{-k(t-5)}$ to obtain the ion loss rate without MOT ($k_1$) and with MOT ($k_2$) for each of the nine data sets. The mean values from the nine experimental runs are: $k_1 = 0.089$ (0.014) s$^{-1}$ and $k_2 = 0.056$ (0.018) s$^{-1}$, where the values in the parentheses are the standard deviation (s.d.) of means. The corresponding lifetimes without MOT and with MOT are $\tau_1 = 11.2^{+2.2}_{-1.6}$ s and $\tau_2 = 17.8^{+8.2}_{-4.3}$ s, respectively. The increase in ion lifetime is due to cooling of $^{39}$K$^+$ ions by the ultracold $^{85}$Rb atoms in the MOT.

In Fig. 2(b) we plot, for $t \geq 5$ s, the ratio between the number of trapped $^{39}$K$^+$ ions in presence and absence of the $^{85}$Rb MOT, obtained by averaging over the nine experimental runs. The ratio increases with increasing $t$ and a fit to an expression of the form $A\, e^{k_{eff}(t-5)}$ yields $A = 1.02\ (\pm 0.04)$ and $k_{eff} = 0.0293\ (\pm 0.0024)$ s$^{-1}$. The agreement between $k_{eff}$ and $(k_1 - k_2)$ provides a consistency check on the fits and ascertains the observed increase in ion lifetime. The ion temperature without and with MOT increase approximately at rates $\approx k_1 U$ and $\approx k_2 U$, respectively, and thus the estimated net cooling rate ($R$) due to the presence of the MOT is $R \approx k_{eff}\, U \approx 14.7 \pm 1.2$ meV/s (see SM [27] for an ion cooling model).

In addition to the $^{85}$Rb MOT, there is the presence of the background gas (b.g.) of Rb, trace amounts of K and other unidentified gases in our chamber. We observe that the K$^+$ ion cooling is suppressed [Fig. 2(b)], when performing the same experiment at elevated b.g. pressure (obtained by operating the Rb dispenser at higher currents), while adjusting the MOT parameters such that the atom numbers and density are approximately the same. The Rb background vapour contributes to the heating of the $^{39}$K$^+$ ions consistent with ion heating by uniform buffer gas atoms of higher mass as predicted initially by Major and Dehmelt [15] and recently by others [16–18].

We note that in earlier experiments with $^{85}$Rb$^+$ and $^{85}$Rb [12,24], the cooling of ions was faster because the ions and atoms were of the same mass for which collisional cooling is more efficient and, in addition, resonant charge exchange (RCE) might also have been active [29]. Since K and Rb are different species RCE is not possible, but non-resonant charge exchange (nRCE) is. However, careful experimental tests allow us to conclude that this channel is too weak [30] and is not detected in our experiment (see SM [27]). Therefore the dominant cooling channel is multiple elastic collisions, which reduces the cooling rate of the $^{39}$K$^+$ ions by the $^{85}$Rb MOT as compared to cooling of $^{85}$Rb$^+$ by $^{85}$Rb in earlier experiments [12,24]. The essence of the present experiment is that in the ideal situation with no background gas, a spatially small MOT at the precise centre of the ion trap would always cool a trapped ion via elastic collisions, irrespective of the ion-atom mass ratio but at different rates.

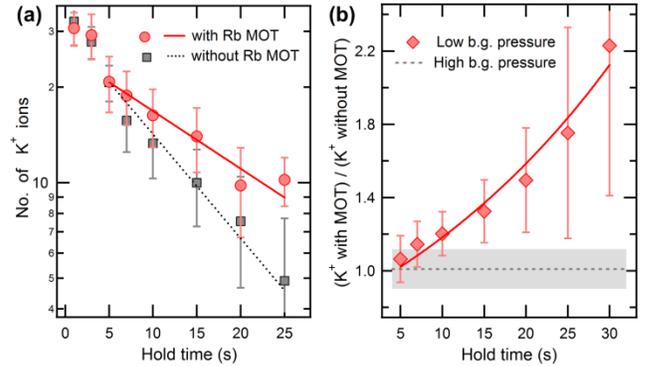

FIG. 2 (Color online). (a) The number of $^{39}$K$^+$ ions remaining in the ion trap for different values of hold time either in absence (squares) or presence (circles) of the $^{85}$Rb MOT. The dotted and the solid lines are single exponential fits for the respective cases (see text). The increase in survival probability of trapped $^{39}$K$^+$ ions in the presence of the $^{85}$Rb MOT indicates cooling of $^{39}$K$^+$ ions. The error bars represent the width (1 s.d.) of the underlying ion number distribution. (b) The ratio between the number of trapped $^{39}$K$^+$ ions in presence and absence of the $^{85}$Rb MOT, for low ($\leq 5.9\times10^{-10}$ Torr) background partial pressure of Rb (symbols) and for high ($\geq 8.3\times10^{-10}$ Torr) background partial pressure of Rb (dotted line). The shaded region represents 1 s.d. for the measurements at high background pressure. At high background pressure, the cooling of $^{39}$K$^+$ ions (indicated by ratios >1), is not experimentally discernible. The solid line is a fit to an exponential for the low background pressure case (see text). The quoted partial pressure of Rb was obtained from the loading rate of the $^{85}$Rb MOT [28].

*Rubidium ion cooling by caesium atoms.* To further validate our results, we perform another experiment in an entirely different experimental apparatus consisting of a linear Paul trap for $^{85}$Rb$^+$ ions [12,21] and a MOT for ultracold $^{133}$Cs atoms (see SM [27]). To demonstrate cooling of $^{85}$Rb$^+$ ions by $^{133}$Cs atoms, we follow an experimental sequence similar to Fig. 1(d) except that the first, fourth and fifth rows of the figure now are $^{85}$Rb laser, $^{133}$Cs laser and $^{133}$Cs MOT, respectively. The result of the experiment is shown in Fig. 3 – a larger number of $^{85}$Rb$^+$ ions survive when held in the presence of the $^{133}$Cs MOT. This can only be explained by the cooling of $^{85}$Rb$^+$ ions due to ultracold $^{133}$Cs atoms. The inset in Fig. 3 shows the ratio between the number of surviving $^{85}$Rb$^+$ ions in the presence and the absence of the $^{133}$Cs MOT, obtained by averaging over ten experimental runs. The ratio increases with increasing $t$ and a fit to an expression of the form $A\,e^{k_{eff}(t-1)}$ yields $A = 1.00$ ($\pm 0.03$) and $k_{eff} = 0.053$ ($\pm 0.003$) s$^{-1}$. This value of $k_{eff}$ is much greater than that for cooling of $^{39}$K$^+$ ions by $^{85}$Rb atoms shown earlier. This is partly because the atom-ion mass ratio (1.565) in this case is comparatively lower, resulting in faster cooling compared to the $^{39}$K$^+$ ion – $^{85}$Rb atom case. Notably, for long hold times, a significant reduction in the width of the $^{85}$Rb$^+$ ion arrival time distribution is observed when the $^{133}$Cs MOT is present [e.g. at 13 s, (width with MOT)/(width without MOT) = 0.91 ± 0.01], providing independent evidence for the cooling of $^{85}$Rb$^+$ ions by the $^{133}$Cs atoms. Further, since this experiment is done in a different kind of ion trap, the ion cooling seems to be robust against changes in ion trap parameters. For brevity, in the rest of the manuscript we focus our discussion on cooling of $^{39}$K$^+$ ions by $^{85}$Rb atoms.

*Simulations.* We use a MC algorithm, and solve the equations of motion to track how the ions collisionally cool (see SM [27]). In Fig. 4(a) we plot the lifetime of 30 non-interacting $^{39}$K$^+$ ions in the presence of a $^{85}$Rb MOT of FWHM $d = 235$ μm, whose atoms are at zero velocity, and a background gas (b.g.) of Rb, whose velocity distribution is consistent with room temperature. As the density of the b.g. decreases with respect to the MOT peak density, the effective cooling rate increases, and ion survives longer. When the b.g. density is zero, the ions are held in perpetuity. Figure 4(b) illustrates the competition between the cooling and the heating of the ion, for the cases in Fig. 4(a) and shows cooling to a steady state temperature when background vapour of Rb is absent. In absence of b.g., the steady state temperature $T_s$ of the ions is proportional to $d^2$ for given trap parameters. In the SM [27], we discuss the simulations in greater details and show that ions are cooled with localized and precisely centred ensembles of atoms irrespective of the atom-ion mass ratio.

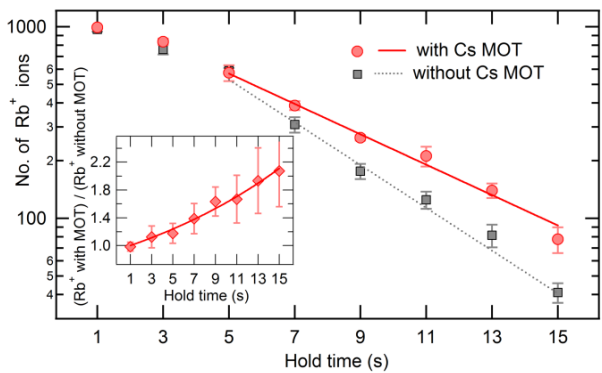

FIG. 3 (Color online). The number of $^{85}$Rb$^+$ ions remaining in the linear Paul trap for different values of hold time either in absence (squares) or presence (circles) of the $^{133}$Cs MOT. The dotted and the solid lines are single exponential fits for the respective cases. The error bars represent the width (1 s.d.) of the underlying ion number distribution. Inset: The ratio between the number of surviving $^{85}$Rb$^+$ ions in presence and absence of the $^{133}$Cs MOT.

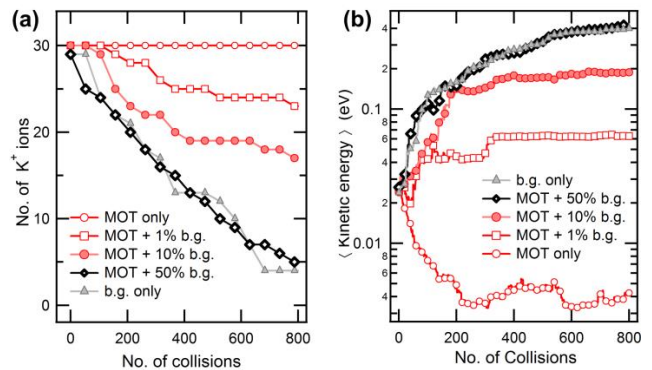

FIG. 4 (Color online). (a) The number of surviving $^{39}$K$^+$ ions in the simulation as a function of the number of collisions each ion experiences. The ions collide with both the background gas of Rb atoms and the $^{85}$Rb MOT atoms. At high background pressures, the ion losses with and without MOT are the same, as seen experimentally in Fig. 2(b). (b) The mean kinetic energy of the ions for the above simulations. As the background gas density is reduced, cooling due to the MOT overcomes the heating due to background vapour and therefore lowers the ion kinetic energy.

*Conclusion.* The experiments demonstrate collisional cooling of light ions by heavier atoms. The reported result is also a demonstration in support of no critical mass ratio for the cooling of ions with atoms, in hybrid traps. Judicious choice of ion-atom species, MOT spatial extent and the secular trap depth of the ion trap could allow a single trapped ion to be cooled to very low temperatures, where possibly the low partial wave regime of ion-atom collisions can be attained at least momentarily [31,32]. Our demonstration also opens up the possibility of simultaneously trapping multiple ionic species with an ensemble of ultracold atomic gas, irrespective of the mass ratio with applications in cold ion chemistry and laboratory astrophysics [33] with trapped protons, $H_2^+$ or $HD^+$ collisionally cooled with $^6Li$.

*Acknowledgments.* We thank T. Ray, S. Jyothi, A. Sharma, K. Ravi and S. Lee for the development of the experimental apparatus and techniques; M. Ibrahim for developing Fig. 1(a). The MC collision codes are modified versions of the codes created by K. Ravi and S. Lee. S.D. acknowledges support from the Department of Science and Technology (DST) in the form of the DST-INSPIRE Faculty Award (IFA14-PH-114).

**Supplementary Material:**

*Magneto-optical trap (MOT) for $^{85}Rb$ and $^{39}K$.* The $^{85}Rb$ MOT is formed by three pairs of mutually orthogonal laser beams, each pair consisting of two counter propagating beams. Each of these beams have light at two frequencies – the cooling light that is 12 MHz red detuned from the $^{85}Rb$ ($5S_{1/2}$, $F = 3$) → $^{85}Rb$ ($5P_{3/2}$, $F' = 4$) transition and the repumping light that is on resonance with the $^{85}Rb$ ($5S_{1/2}$, $F = 2$) → $^{85}Rb$ ($5P_{3/2}$, $F' = 3$) transition. The total cooling (repumping) beam power is 20 mW (1 mW) distributed equally in the six beams. With a magnetic field gradient of ~22 Gauss/cm a $^{85}Rb$ MOT with ~ $4\times10^5$ ultracold $^{85}Rb$ atoms, FWHM ~ 280 μm and density ~ $10^{10}$ cm$^{-3}$ is formed at a typical temperature of 150 μK. The $^{39}K^+$ ions are created by photoionizing ultracold $^{39}K$ atoms from a $^{39}K$ MOT. The cooling beam for the $^{39}K$ MOT is ~35 MHz red detuned from the $^{39}K$ ($4S_{1/2}$, $F = 2$) → $^{39}K$ ($4P_{3/2}$, $F' = 3$) transition and the repumping beam is ~15 MHz red detuned from the $^{39}K$ ($4S_{1/2}$, $F = 1$) → $^{39}K$ ($4P_{3/2}$, $F' = 2$) transition.

The beams for the $^{39}$K MOT are combined with those for the $^{85}$Rb MOT on a beam splitter and then propagate along the same direction. The two MOTs are formed at the same location. The loading of the two MOTs are independently controlled by mechanical shutters placed in the paths of the respective laser beams.

*$^{39}$K$^+$ ion trapping and detection.* The optimized ion trap operation for $^{39}$K$^+$ ions involves applying a sinusoidal rf voltage $V_{rf}$ of amplitude 85 V and frequency 700 kHz to the inner pair of wires (spaced by 3 mm) while the outer pair (spaced by 6 mm) is biased at -5 V [see Fig. 1(a) in the main manuscript]. At this setting, the chances of trapping $^{85}$Rb$^+$ ions, created for example by non-resonant charge exchange (nRCE), is greatly reduced but still probable. The number of trapped ions is detected by applying a high voltage (360 V) pulse that extracts the ions from the ion trap and launches them along the axis of a time of flight (ToF) mass spectrometer (MS) at the end of which is the channel electron multiplier (CEM). The ToF MS enables differentiation between $^{85}$Rb$^+$ and $^{39}$K$^+$ ions based on their arrival time. The above setting was used for the data reported in Fig. 2 of the manuscript.

*Precise alignment of the $^{85}$Rb MOT with the center of the $^{39}$K$^+$ ion trap.* We have to align the $^{85}$Rb MOT with the ion trap center very carefully to observe the cooling of $^{39}$K$^+$ ions. In our apparatus, this alignment is facilitated by an optical cavity whose center, by construction, is well matched with the ion trap center [S1,S2]. The $^{85}$Rb MOT is aligned with the cavity and the ion trap center by (*i*) optimizing the $^{85}$Rb MOT fluorescence coupled out of the cavity and (*ii*) by measuring the vacuum Rabi splitting of the coupled atom-cavity system [S3]. We note that the cooling of $^{39}$K$^+$ ions is completely suppressed on shifting the $^{85}$Rb MOT away (even by a few 100 μm) from the center of the ion trap.

*Checks for non-Resonant Charge Exchange (nRCE).* In order to establish that nRCE (i.e. $^{39}$K$^+$ + $^{85}$Rb → $^{39}$K + $^{85}$Rb$^+$) is not active in the present experiment we adopted the following protocol. (*a*) In an independent measurement, $^{85}$Rb$^+$ ions are loaded into the ion trap by photoionization of $^{85}$Rb atoms from a $^{85}$Rb MOT. The $^{85}$Rb$^+$ ions are held in the ion trap with or without $^{85}$Rb MOT for a certain hold time and then exacted onto the CEM for detection. The arrival time distribution of $^{85}$Rb$^+$ ions is recorded. (*b*) Keeping the ion trap voltages identical, $^{39}$K$^+$ ions are loaded into the ion trap by photoionization of ultracold $^{39}$K atoms in a $^{39}$K MOT. The $^{39}$K$^+$ ions are held in the ion trap for a certain hold time (without the $^{85}$Rb MOT, thus precluding $^{85}$Rb$^+$ ion formation) and then exacted onto the CEM for detection. The arrival time distribution of $^{39}$K$^+$ ions is recorded and found to be distinct from $^{85}$Rb$^+$ ions. (*c*) The experiment with $^{39}$K$^+$ ions in the $^{85}$Rb MOT is then performed, and the ion arrival time distribution is found to be consistent with $^{39}$K$^+$ ions but not with $^{85}$Rb$^+$ ions. No confirmed detection of $^{85}$Rb$^+$ ions could be made suggesting low rates of nRCE.

We also perform another experiment to verify that charge exchange can be ruled out. In this experiment we operate the ion trap at reduced amplitude of the rf voltage (69 V). At this setting $^{85}$Rb$^+$ cannot be trapped at all but $^{39}$K$^+$ ions can still be trapped efficiently. In the experiment, this implies that if $^{85}$Rb$^+$ were formed by nRCE they would leave the trap and the number of ions detected would decrease when $^{39}$K$^+$ ions are held in presence of the $^{85}$Rb MOT. In contrast, however, an increase in the number of ions was observed conforming that nRCE rates are low. The cooling of $^{39}$K$^+$ ions by $^{85}$Rb atoms was observed even at this altered ion trap setting which allows us to confidently conclude that nRCE is negligible and the ion cooling effect is robust against changes in ion trap parameters.

*Experiment for cooling of $^{85}$Rb$^+$ ions by $^{133}$Cs atoms.* The experiment for cooling of $^{85}$Rb$^+$ ions by $^{133}$Cs atoms is performed in a different apparatus. As shown schematically in Fig. 1S, it consists of a linear Paul trap for $^{85}$Rb$^+$ ions [S4,S5] and magneto-optical traps for $^{85}$Rb and $^{133}$Cs atoms. The set up for the $^{85}$Rb MOT is similar to the one described above, and hence is not described again. The $^{133}$Cs MOT is loaded from a Cs dispenser source. The $^{133}$Cs MOT is formed by three mutually orthogonal beams, each of which is retro-reflected onto itself. The beams consist of light at both the cooling and repumping frequencies. The cooling light is ~15 MHz red detuned from the $^{133}$Cs ($6S_{1/2}$, $F = 4$) → $^{133}$Cs ($6P_{3/2}$, $F' = 5$) transition and the repumping light is on resonance with the $^{133}$Cs ($6S_{1/2}$, $F = 3$) → $^{133}$Cs ($6P_{3/2}$, $F' = 4$) transition. The total cooling (repumping) beam power is 15 mW (1.5 mW) distributed equally in the beams. With a magnetic field gradient of ~17 Gauss/cm a $^{133}$Cs MOT with ~ $10^6$ ultracold $^{133}$Cs atoms and FWHM ~ 550 μm is formed.

In the ion trap, the radial confinement is provided by a sinusoidal voltage of amplitude 85 V and frequency 500 kHz applied to the rod electrodes such that voltages on the adjacent rods have 180° phase difference. Confinement in axial direction is provided by a DC voltage of 80 V applied to the ring end cap electrodes. To detect the trapped ions, the voltage on one of the ring end cap electrodes is changed from 80 V to -3 V. This extracts the ions from the trapping region and launches them towards the CEM for detection. The $^{85}$Rb$^+$ ions required for the experiment are created by resonant two photon ionization of $^{85}$Rb atoms in the $^{85}$Rb MOT, where the first photon (cooling laser) results in the $5S_{1/2} \rightarrow 5P_{3/2}$ excitation at 780 nm and a second photon is sourced from a blue (473 nm) laser beam.

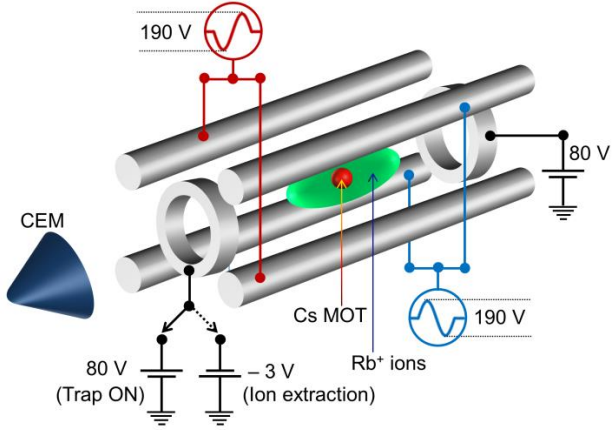

FIG. 1S. A schematic representation of the apparatus for trapping of $^{85}$Rb$^+$ ions (in a linear Paul trap) and ultracold $^{133}$Cs atoms (in a MOT). The rf voltages for radial confinement of ions is applied to the four rod electrodes and the DC voltages for axial confinement is applied to the two ring end cap electrode. The extraction of the trapped ions to the CEM is done by switching the voltage on one of the ring end cap electrodes from 80 V to -3 V.

To measures the lifetime of the trapped $^{85}$Rb$^+$ ions when held with and without the $^{133}$Cs MOT atoms, an experimental sequence similar to Fig. 1(d) is used. As the blue light can also ionize $^{133}$Cs MOT atoms in a similar two photon process, the MOTs are operated sequentially to prevent the ionization of $^{133}$Cs atoms. First, the $^{85}$Rb MOT is loaded, ion trapping fields are switched on, $^{85}$Rb$^+$ ions are created by switching on the blue laser briefly, and the $^{85}$Rb MOT light is shuttered off which empties the $^{85}$Rb MOT. Subsequently, either the $^{133}$Cs MOT light is allowed in and the $^{133}$Cs MOT is loaded or the $^{133}$Cs light is kept blocked and no MOT is loaded. The trapped $^{85}$Rb$^+$ ions are held for variable hold times and extracted for detection by the CEM. The surviving ion number in the ion trap is determined as a function of ion hold time, in the presence or absence of the $^{133}$Cs MOT. A representative plot from the experiment is provided in Fig. 3 of the manuscript. An increase in lifetime of trapped $^{85}$Rb$^+$ ions due the presence of $^{133}$Cs MOT is observed. This implies that trapped $^{85}$Rb$^+$ ions are cooled by heavier, localized, ultracold $^{133}$Cs atoms. We note that we checked for nRCE (i.e. $^{85}$Rb$^+$ + $^{133}$Cs $\rightarrow$ $^{85}$Rb + $^{133}$Cs$^+$) following the protocol discussed above but did not detect any. For brevity, in the rest of this document we will focus on cooling of K$^+$ ions by Rb atoms.

*Setting up the K$^+$ ion cooling model.* The trapped K$^+$ ion collides with Rb atoms contained in two distinct thermal baths. The background Rb vapour at room temperature, which floods the entire vacuum chamber is the first bath, called background gas bath (b.g.b.), and the cold MOT of $^{85}$Rb atoms, localized within a small volume at the centre of the ion trap is the MOT bath (m.b.). Since the MOT loads from the background vapour, the experiments with MOT are necessarily performed in the presence of both the baths. The survival of the ions in the ion trap is determined by the exchange of energy between the ions and the thermal baths. The number of trapped ions decreases with time in both cases but at different rates, which allows the explicit determination of the cooling rate of the trapped ions due to the $^{85}$Rb MOT atoms, in the model described below. To keep the discussion concise and specific, we will apply the model to obtain the cooling rate for the data presented in Fig. 2(a) of the manuscript.

*Ion cooling model.* For the model, the ion trap with trap depth of $U \approx 0.5$ eV and radial and axial secular frequencies $\Omega_r/2\pi$ ($\approx 66$ kHz) and $\Omega_z/2\pi$ ($\approx 100$ kHz) is treated as a three-dimensional isotropic ion trap with the secular frequency $\Omega_s/2\pi \approx 80$ kHz. The $^{39}$K$^+$ ions are created from a MOT of $^{39}$K atoms, with $\sigma \sim 330\ \mu m$, which defines the initial radial distribution of ions ( $\frac{4\pi N_i}{(\sigma\sqrt{2\pi})^3} r^2 \exp\left[-\frac{r^2}{2\sigma^2}\right]$, where $N_i = 34$ and $\sigma = 330\ \mu$m). Given the initial ion distribution, the secular potential trap depth, secular frequency and the experimental rate of ejection from the ion trap, the rate at which the ion distribution energy increases with time is completely constrained, and can therefore be calculated. Since for both b.g.b. and (b.g.b. + m.b.) cases, the ions are heating out of the trap, the same model can be used for determining the rate of change of energy of the ion distribution.

To simplify the analysis, we ignore that the ion decay is slightly faster for $t < 5$ s (Fig. 2a) and fit the ion number without MOT to a single exponential decay function ($N_i e^{-t/\tau}$) starting from $t = 0$. The data with MOT is fit to a single exponential decay for $t \geq 5$ s. The rate of increase in the ion distribution energy with time can be deterministically calculated by transforming the ion distribution so that the correct ion number is present in the trap at any instant of time, for both cases. This is done by partitioning the radial extent of the ion trap into concentric zones, where in each time interval (1 second), ions (from the distribution) in the n$^{th}$ radial zone shift into the (n+1)$^{th}$ radial zone, thus increasing their total energy. The radial zones differ for the cases of b.g.b. and (b.g.b. + m.b.) in a way that is consistent with the measured ion ejection rate for the respective cases. Ions in the outermost zone of the trap escape and do not gain further energy. In the limit of large time, when all the ions have escaped from the trap, the energy per ion ($E_I$) reaches its maximum value ($U$). Fixing the ion decay rates to the fitted values, we can then explicitly calculate the increase in $E_I$, and the mean energy of all the ions $\langle E_I \rangle = \sum_{j=1}^{N_i}(E_I)_j/N_i$ at any time. The result of this calculation is shown in Fig. 2S, where the squares show $\langle E_I \rangle$ vs. time for the background gas of Rb only (b.g.b.) and the circles show $\langle E_I \rangle$ vs. time in the presence of the MOT (b.g.b. + m.b.). As with the experiment, since the $^{85}$Rb MOT loads for ~ 5 seconds after the creation of the $^{39}$K$^+$ ions, the cooling effect due to the presence of the MOT is evident only from 5 seconds onward. The solid lines through the points in Fig. 2S are not fits to the points, but the function

$$\langle E_I(t)\rangle = \langle E_I^0 \rangle + (U - \langle E_I^0 \rangle)(1 - exp[-t/\tau])$$

where $\tau$ denotes either $\tau_1 = 11.2$ s (for the without MOT case) or $\tau_2 = 24.6$ s (for the with *Rb* MOT case) from the fits to one of the better data sets [shown in Fig 2(a)] among the nine data sets. $\langle E_I^0 \rangle$ is the initial mean energy per ion for the respective cases. The dotted lines around the two solid curves represents $\langle E_I(t)\rangle$ for one standard deviation in the time constants $\tau_1$ and $\tau_2$. The difference in measured trap lifetimes for the two distinct cases b.g.b. and (b.g.b. + m.b.), translates into the energetics of the trapped ions. The slopes of the curves in Fig. 2S give the rate of change in mean ion energy: $\left[\frac{d\langle E_I(t)\rangle}{dt}\right]_{t=5}^{b.g.b.} = 26.2^{+1.2}_{-1.2}$ meV/s for the case of only background gas vapour and $\left[\frac{d\langle E_I(t)\rangle}{dt}\right]_{t=5}^{b.g.b.+m.b.} = 12.4^{+1.4}_{-1.1}$ meV/s in the presence of the $^{85}$Rb MOT. The difference between the two heating rates gives the ion cooling rate $R$ at $t = 5$ s, due to the presence of the $^{85}$Rb MOT as $R = 13.8$ meV/s. The cooling rate is completely consistent with the rate (14.7 ±1.2 meV/s) estimated in the main text of the manuscript.

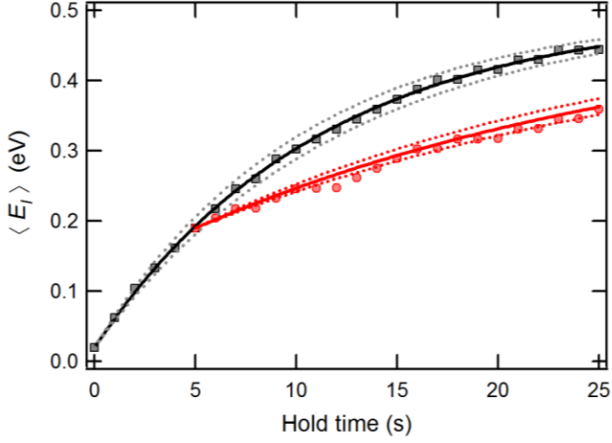

FIG. 2S. The change in the mean energy of the ions with time is shown. The squares show the variation in ion energy for the b.g.b. case while the circles illustrate this for the (b.g.b. + m.b.) case. Clearly the MOT has a cooling effect, the magnitude of which is quantified in the text. The solid lines through the points represent the exponential rise function (see text) with $\tau_1 = 11.2$ s and $\tau_2 = 24.6$ s for b.g.b. and (b.g.b. + m.b.) case, respectively, and the dotted lines around each data set represent the same function when $\tau_1$ and with $\tau_2$ deviate by one standard deviation from their mean values, respectively. The effect of the cooling of ions due to the MOT is clearly seen. The slopes of the curves give the rate of change of the ion energy for the two cases and therefore the cooling rate due to the MOT, as discussed in the text.

*Simulation methods.* In order to study the effect of collisions on the trajectory, kinetic energy and the survival probability of an ion in the ion trap we use Monte Carlo method to simulate, in Mathematica, the dynamics of the ion. The static electric field profile of the modified spherical Paul trap is calculated in SIMION [S1,S2]. This field profile is imported into Mathematica, converted into a time dependent potential by appropriate use of the superposition principle, and used in the Monte Carlo simulation. In the simulations the ions are taken as non-interacting, which is a good approximation in the regime of ion temperatures and the small ion numbers in our experiment. (The small number and high temperatures of ions ensures that the likelihood of two ions being present in the trap centre at the same time is very small because ions move very quickly through the trap centre. In the outer reaches of the ion trap, the ion-ion collisions are neglected as the volume of the trap makes ion-ion collisions in the outer reaches of the trap very rare.) We evolve the trajectory of a single ion in the ion trap and introduce instantaneous collisions with atoms in between successive evolutions of the ion. The code is a faithful representation of the conceptual framework for multiple scattering of ions by atoms, and is adapted from the simulations in Ravi *et al.* [S5] with two major differences: (*i*) the charge exchange collisions are dropped i.e. only elastic collisions are considered, and (*ii*) collision with a uniform density background gas is also included.

At the start of the simulation the ion is stationary, which represents the creation of an ion from a MOT with negligible initial kinetic energy, and its position is chosen randomly within a sphere of radius 1 mm that is co-centric with the ion trap. To incorporate the effect of collisions with MOT atoms we introduce a collision probability that has the same spatial profile as the atomic density profile, namely a Gaussian density distribution with FWHM $d = 0.235$ mm, co-centred with the ion trap. In addition, the effect of background atomic gas (Rb as in the experiment) is incorporated by allowing for collision everywhere within the ion trap. The probability of colliding with background gas, as opposed to colliding with the MOT atoms, depends on the relative density of the background gas with respect to the local density of the MOT. When the ion collides with a background gas atom, the velocity of the colliding atom is randomly chosen from a Maxwell-Boltzmann distribution at the temperature of 300 K. The velocity of the atom in the MOT is taken to be zero (since the velocity of a cold atom is much lower than the typical velocities of the ions). The time when it is decided to have a collision is taken from a Poisson distribution with a mean corresponding to 20 rf cycles which is greater than the time period of the ion's secular motion (the largest time scale of ion trap). The ion is then evolved for 20 rf cycles and the position of collision is chosen randomly with the weight of the *total* density distribution of the atoms. After each collision, the ion's evolution in the trap starts from the same position but with a different velocity determined by the kinematics of the collision [S5,S6]. We monitor the kinetic energy (K.E.) of the ion till the time it remains in the ion trap. The ion temperature is recovered from the mean K.E. and the equipartition theorem. We evolve 30 such instances (which corresponds well to the total number of ions loaded in the ion trap in the $^{85}$Rb atom - $^{39}$K$^+$ ion experiment) to observe the average effect with different initial conditions. The mean K.E. between two successive collisions is averaged over the 30 instances to monitor the effects of collisions on the ion temperature. In the simulation, the position of the ion is also monitored. If the ion goes beyond the ion trap region, the ion evolution is halted and the ion is lost. The trap extends approximately 4 mm in x, y and z directions.

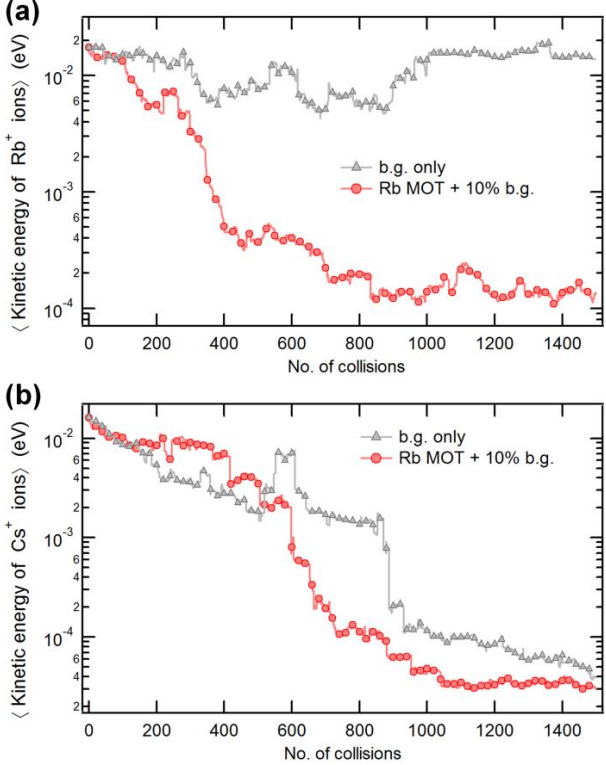

FIG. 3S. Collisional cooling of ions for different ion-atom mass ratios. (a) The mean kinetic energy of trapped $^{85}$Rb$^+$ ions as a function of number of collisions with equal mass $^{85}$Rb atoms. Triangles represent the case when collision with background gas (b.g.) of Rb alone is considered. Circles represent the case when collisions with both $^{85}$Rb MOT atoms and b.g. are considered. No net change in mean K.E. is seen when b.g. alone is considered, consistent with the original theory by Major and Dehmelt [S7] for $m_a = m_i$. The addition of the $^{85}$Rb MOT, i.e. a localized atomic density at the center of the ion trap, results in a decrease of the mean K.E. suggesting a decrease in the temperature of trapped $^{85}$Rb$^+$ ions. (b) The mean K.E. of trapped $^{133}$Cs$^+$ ions as a function of number of collisions with lighter $^{85}$Rb atoms. Triangles represent the case when collision with background gas (b.g.) of Rb alone is considered. Circles represent the case when collisions with both $^{85}$Rb MOT atoms and b.g. of Rb are considered. A net decrease in the mean K.E. is seen in both cases. The case with b.g. only is consistent with the original theory by Major and Dehmelt [S7] for $m_a < m_i$. The addition of the $^{85}$Rb MOT increases the rate at which the mean K.E. decreases, suggesting a faster rate of cooling of the trapped $^{133}$Cs$^+$ ions and a lower final temperature. It is noteworthy that in both cases (a) and (b) above, the simulation also shows that a trapped ion never escapes the ion trap.

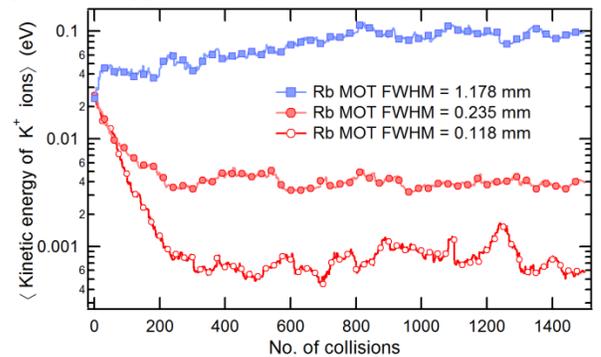

FIG. 4S. Effect of MOT size on the temperature of trapped $^{39}$K$^+$ ions. The mean K.E. of trapped $^{39}$K$^+$ ions as a function of the number of collisions with heavier $^{85}$Rb MOT atoms is shown for three different sizes of the $^{85}$Rb MOT. Background gas collisions are completely neglected. Beyond a critical MOT size, a net increase in the mean K.E. is seen (filled squares). This is explained since the limiting case, that of infinite MOT size, is expected to be consistent with Major and Dehmelt's theory [S7] of ion heating when $m_a > m_i$. Smaller MOTs, shown as filled and open circles, enable faster rate of decreases of mean K.E. and result in lower final temperatures of the trapped ions. The steady state temperature $T_s$ of the ions is found to be approximately proportional to the square of $d$ and square of $\Omega_s$, for a given ion-atom pair.

The above mentioned simulation was done with three different ions: $^{39}$K$^+$, $^{85}$Rb$^+$ and $^{133}$Cs$^+$ in a gas of $^{85}$Rb atoms. These three cases correspond to different atom-ion mass ratios ($m_a/m_i$). The ratios are = 2.179, 1 and 0.639, corresponding to low, equal and high mass of the ion with respect to the colliding atom. In all these cases, the $^{85}$Rb MOT is assumed to have a Gaussian density distribution with FWHM = 0.235 mm and the relative probability of colliding with background gas (b.g.) is 0.1 (i.e. 10%). In Fig. 3S, we show the results for $^{85}$Rb$^+$ and $^{133}$Cs$^+$ ions. The results for $^{39}$K$^+$ ions are shown in Fig. 4S and in Fig. 4 of the main manuscript.

———————————